\documentclass[preprint,12pt]{elsarticle}
\usepackage{color,soul}
\usepackage{graphicx}
\usepackage{algpseudocode}
\usepackage{setspace}
\usepackage{booktabs}
\usepackage[font=small,skip=0pt]{caption}
\usepackage{amsthm}
\usepackage{hyperref}
\usepackage{amssymb}
\usepackage{commath}
\usepackage[ruled, vlined, linesnumbered]{algorithm2e}
\usepackage{bm}

\makeatletter
\newcommand{\LeftEqNo}{\let\veqno\@@leqno}
\makeatother

\journal{Journal of Biomedical Informatics} 	
\begin{document}
\begin{frontmatter}

\title{CAMIRADA: Cancer microRNA association discovery algorithm, a case study on breast cancer}

\author[add1]{Sepideh Shamsizadeh\corref{cor1}}
\ead{s.shamsizadeh@ut.ac.ir}

\author[add1]{Sama Goliaei}
\ead{sgoliaei@ut.ac.ir}

\author[add1,add2]{Zahra Razaghi Moghadam}
\ead{razzaghi@ut.ac.ir}

\cortext[cor1]{Corresponding author}
\address[add1]{Faculty of New Sciences and Technologies, University of Tehran, Tehran, Iran}
\address[add2]{Max Planck Institute of Molecular Plant Physiology, Posdam, German}

\begin{abstract}In recent studies, non-coding protein RNAs have been identified as microRNA that can be used as biomarkers for early diagnosis and treatment of cancer, that decrease mortality in cancer. A microRNA may target hundreds or thousands of genes and a gene may regulate several microRNAs, so determining which microRNA is associated with which cancer is a big challenge. Many computational methods have been performed to detect micoRNAs association with cancer, but more effort is needed with higher accuracy. Increasing research has shown that relationship between microRNAs and TFs play a significant role in the diagnosis of cancer. Therefore, we developed a new computational framework (CAMIRADA) to identify cancer-related microRNAs based on the relationship between microRNAs and disease genes (DG) in the protein network, the functional relationships between microRNAs and Transcription Factors (TF) on the co-expression network, and the relationship between microRNAs and the Differential Expression Gene (DEG) on co-expression network. The CAMIRADA was applied to assess breast cancer data from two HMDD and miR2Disease databases. In this study, the AUC for the 65 microRNAs of the top of the list was 0.95, which was more accurate than the similar methods used to detect microRNAs associated with the cancer artery.
\end{abstract}


\end{frontmatter}

\section{Introduction} \label{I}

According to a study,breast cancer is one of the most common causes of women's deaths in cancer \citep{yang2015crucial}. The molecular traits of the primary tumors play an important role in timely diagnosis and treatment of the next stages of breast cancer. Therefore, we need methods that can detect cancer at an early stage in order to help clinicians and patients to be treated with biological markers \citep{yang2015crucial}.
In recent years, many studies have been conducted to clarify the mechanisms that make cancer progress and develop.

Although many genes that cause cancer and stop it are known by researchers, it is still necessary to identify cancer pathways. Therefore, one of the most important biological targets of cancer is the detection of genes related to cancers \citep{farahmand2016gta,razaghi2016hybridranker}. 

MicroRNAs are short non-coding RNAs with an approximate length of 22 nucleotides that are involved in post-transcriptional regulation and that are major regulators of gene expression \citep{guruceaga2014functional}.In 2002, the first link between cancer and microRNAs was identified \citep{adali2012analysis}. MicroRNAs target about 60\% of human genes that be involved in a wide range of biological processes and disease including cell division, proliferation, differentiation, and apoptosis \citep{le2015network}. More than 50\% of human microRNAs are located in cancer-associated genomic regions. MicroRNAs could play roles as oncogenes or tumor suppressor genes. Dysregulation of microRNAs expression has been shown to have impacts on human diseases \citep{le2015network}. 

However, identifying the related microRNAs with existing experimental tests may be difficult and time consuming. In addition, many researchers are also faced with limited knowledge about microRNAs. Therefore, a number of computational approaches have been recently developed to identify microRNAs associated with the disease. Statistical methods, machine learning, and network-based methods are some of the methods proposed to predict cancer-related microRNAs.

Zhao et al. developed a framework for obtaining the relationship between cancer and microRNAs through their target gene expression profiles without requiring neither microRNA expression data or the matched gene and microRNAs expression data. For each microRNA, target genes that are likely to be related to cancer are determined, and subsequently identified adverse pathways associated with cancer. Then, microRNAs are ranked according to these pathways, because high-ranking microRNAs are more likely to be linked to cancer \citep{zhao2014identifying}.

Xue et al. used the Gaussian mixture models, they identified patterns of gene expression for healthy and patient samples, and then, with Fisher's exact test, deduced the regulator relationship between TF and genes. A minimum description length for the pruning of the network, using the (MDL) principle, the relationships between TFs and microRNAs have been achieved \citep{xue2017computational}.

Tseng et al. provided a method to identify active oncomirs and their potential functions in gastric cancer progression. The microRNA and mRNA's expression profiles with the human protein interaction network (PIN) are integrated to show microRNAs-regulated PIN in specific biological conditions. The microRNAs' potential functions were identified by functional enrichment analysis and the activities of microRNAs-regulated PINs were evaluated by the co-expression of protein-protein interactions (PPIs) \citep{tseng2011integrative}.

Le constructed microRNAs functional similarity networks based on shared targets of microRNAs, and then it integrated them with a microRNA functional synergistic network. After analyzing topological properties of these networks, it introduced five network-based ranking methods (RWR, PRINCE (Prioritization and Complex Elucidation), which was proposed for disease gene prediction; PageRank with Priors (PRP) and K-Step Markov (KSM), which were used for studying web networks; and a neighborhood-based algorithm.) in prioritizing candidate microRNAs to predict novel disease-related microRNAs based on the constructed microRNAs functional similarity networks \citep{le2015network}.

In this study, we presented CAMIRADA framework for the identification of microRNAs cancer associations by taking advantage of the useful links between microRNAs targets and disease genes in protein-protein networks (PPIs), the functional connections between microRNAs targets and Transcription Factors(TF) in co-expression network and the functional relationships between microRNAs targets and Differentially Expressed Gene(DEG) in co-expression network.

\section{Methods}\label{II}
\subsection{Data Sets}
We firstly collected the microRNAs target genes predicted with different tools, including PicTar \citep{krek2005combinatorial}, miRanda (version 3.0) \citep{john2004human}, TargetScan (release 6.2) \citep{lewis2005conserved}, miRBase (version 16) \citep{griffiths2006mirbase} and mirTarbase \citep{pinero2016disgenet}. Specifically, for a microRNA, we consider target genes to be obtained at least by three tools. Note that some microRNAs are not considered if, after filtering, its target genes are less than three. In total, we obtained 42832 targeting pairs that involved 825 microRNAs and 8334 target genes.

The disease-gene association data were obtained from DisGeNET \citep{abbott2014candidate} and The Candidate Cancer Gene Database (CCGD) \citep{yang2012chipbase}, which involved 375 genes related to breast cancer. The Transcription factors(TF) were retrieved from ChIPBase \citep{chou2015mirtarbase}, that included 13890 genes. The names of all these genes mapped to Entrez gene id.

\subsection{Human Protein-Protein Interaction (PPI) Data And Co-Expression Network}
PPI data for humans are derived from the Human Protein Reference Database (HPRD Release 9), which contains human-protein annotations based on empirical evidence from published reports \citep{shi2013walking}. We changed the name of the gene to the Entrez gene id, and then we obtained the maximum components of the entire network, which contains 9028 genes and 35865 interactions. It is noteworthy that PPI data in HPRD are the most common protein isoforms, mainly due to lack of experimental data \citep{lewis2005conserved}. 

We downloaded GSE31192 \citep{harvell2013genomic} data from the GEO \citep{barrett2012ncbi} database to build a co-expression network and obtain DEGs. The RNA extracted from breast tissue with microarrays Affymetrix Human Genome U133 Plus 2.0  has been investigated. These gene expression data are pre-processed with the Robust Multichip Average algorithm(RMA). 

Then we find out the Differentially Expressed Genes (DEGs) between normal and abnormal samples with samr package in R. We selected genes, which their $p-value$ were under 0.05 and their $nfold$ was 0.1 . The co-expression network was constructed using the WGCNA \citep{langfelder2008wgcna} package in R software. To choose an appropriate cutoff to include a percentage of the highest correlations, the approximate scale-free topology criterion was applied. For determining the optimal parameter, the function pickSoftThresholdthe in WGCNA package was used. After selecting the best threshold (0.85), adjacency network for co-expression network was calculated.

 The co-expression network included 21179 genes, so for sampling, we considered DEGs and all targets of microRNAs as seed, we mapped them onto the co-expression network, and after that, we applied RWR (Random Walk With Restart)  algorithm for ranking the genes of network \citep{shi2013walking}. Then we selected top 10000 genes and made the network base on their relationships.

\subsection{CAMIRADA algorithm}
\subsubsection{Disease Genes-microRNAs}
We mapped targets of a microRNA and disease genes onto PPI network. After that, we considered disease genes as seeds and applied RWR \citep{shi2013walking} to rank genes of PPI network. Then by using gene set enrichment analysis (GSEA) we derived the ranked gene list, we used ES1 (enrichment score) using the following formula \citep{shi2013walking}.

\begingroup\makeatletter\def\f@size{9}\check@mathfonts
	\begin{equation}\label{eq:01}
		\text{ES1} = \max_{1\le i \le n} \left( \sum_{\substack{g_j \in TG,\\ j\le i} } \sqrt{\frac{N-n_1}{n_1}} - 		\sum_{\substack{g_j \notin TG,\\ j\le i} } \sqrt{\frac{N-n_1}{n_1}}\right)
	\end{equation}
	\endgroup

Where $N$ is the number of all genes in PPI network and $n_1$ is the number of target genes of one microRNA that we showed with TG.  For each microRNA we calculate ES1. We computed a statistical sum from the beginning of the ranked list of PPI's genes. In this way, by moving down the list if the gene was in TG, we added to statistical sum, and if that gene did not exist in the TG, we reduced from statistical sum. The RWR algorithm with each microRNA's target genes as seeds was applied to compute ES2 for the same pairing of microRNAs disease referred to above \citep{shi2013walking}.

\begingroup\makeatletter\def\f@size{9}\check@mathfonts
	\begin{equation}\label{eq:02}
		\text{ES2} = \max_{1\le i \le n} \left( \sum_{\substack{g_j \in DG,\\ j\le i} } \sqrt{\frac{N-n_1}{n_1}} - 	\sum_{\substack{g_j \notin DG,\\ j\le i} } \sqrt{\frac{N-n_1}{n_1}}\right)
	\end{equation}
	\endgroup

Which $n_2$ is the number of disease genes. For all pairs of microRNAs and disease genes we did the above procedures. For computing ES1 we considered disease genes as seed for RWR algorithm and we considered targets of one microRNA as seed for RWR to compute ES2. Then we calculated ES for each pair of microRNAs-disease by using the following formula \citep{shi2013walking}

\begingroup\makeatletter\def\f@size{9}\check@mathfonts
\begin{equation} \label{eq:03}
	\text{ES} = \beta\ \text{ES1} + (1-\beta)\ \text{ES2}
\end{equation}
\endgroup

\subsubsection{DEGs-microRNAs}
We mapped targets of each microRNA and DEG onto co-expression network. For all pairs of microRNAs and DEGs we calculated ES. For computing ES1 we considered DEGs as seed for RWR algorithm and we considered targets of one microRNA as seed for RWR to compute ES2. Then we calculated ES for each pair of microRNAs-DEG Where $N$ is the number of all genes in co-expression network and  $n_1$ is the number of target genes of one microRNA.  For each microRNA we calculate ES1  \ref{eq:01}. Which $n_2$  in the ES2  \ref{eq:02} is the number of DEGs.

\begin{figure}[!t]
		\centering
		\includegraphics[width=.9 \textwidth]{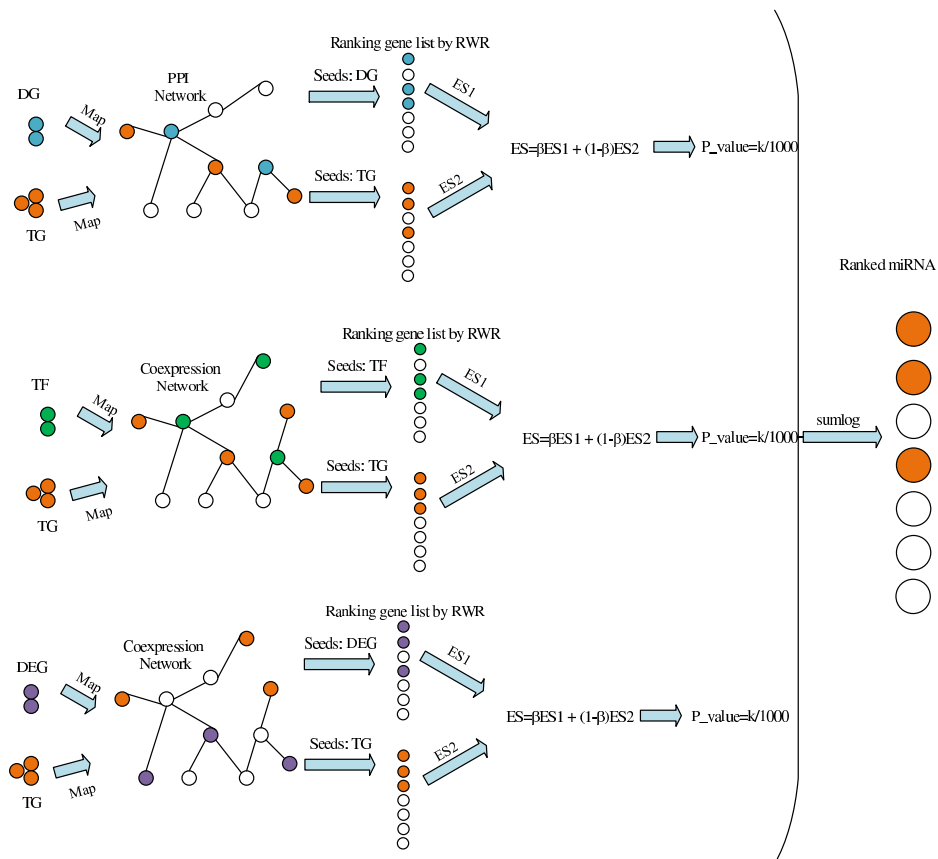}
		\caption{An overview of our framework to predict new diseases associated with microRNAs.}\label{fig:1}		
\end{figure}

\subsubsection{TFs-microRNAs}
We did all above producers on TF and targets of microRNAs. First in order, we mapped targets of all microRNAs and TFs onto co-expression network and for each microRNA we separately saved it's targets in a list. For all pairs of microRNAs and TFs we calculated ES \ref{eq:03}. For computing ES1 we considered TFs as seed for RWR algorithm and we considered targets of one microRNA as seed for RWR to compute ES2. Then we calculated ES for each pair of microRNAs-TFs Where $N$ is the number of all genes in co-expression network and $n_1$ is the number of target genes of one microRNA.  For each microRNA we calculate ES1 \ref{eq:01}. Which $n_2$ in the ES2 \ref{eq:02} is the number of TFs. A review of our frame work are showed in Figure \ref{fig:1}.

\subsection{Random Network And P-value}
To measure the importance of the relationship between microRNAs and disease, we used the p-value. To calculate the p-value, we first created 1000 random networks of the PPI network and 1000 random networks of the co-expression network. Random networks were made in such a way that the number of input and output edges from each node with the number of input and output edges from the same node in the primary network are equal \citep{shi2013walking}. Of course, it should be noted that both the PPI network and the co-expression network are non-directional networks in this study. The p-value is calculated based on \citep{shi2013walking}.

\section{Results And Performance}\label{III}
To evaluate the performance of our algorithm to identify microRNAs-disease associations, we plotted the receiver operating characteristic (ROC) curves and computed the area under the curve (AUC). 

To compute the AUC, we first need to calculate the sensitivity and specificity based on the following formulas, and the sensitivity is defined as the correct detection rate of the positive group on the $y$ axis and the specificity, or the wrong detection rate of the negative category on the $x$ axis. A ROC curve shows a relative compromise between profits and costs \citep{le2015network}.

\begingroup\makeatletter\def\f@size{9}\check@mathfonts
\begin{equation}\label{eq:04}
	1 - \text{specificity} = \frac{FP}{TN+FP}
\end{equation}
\endgroup
\begingroup\makeatletter\def\f@size{9}\check@mathfonts
\begin{equation}\label{eq:05}
	1 - \text{Sensitivity} = \frac{TP}{FN+TP}
\end{equation}
\endgroup

For computing of these we needed to define the set of $TP$، $TN$، $FP$ and $FN$. At first we should define positive set and negative set. For positive set we considered the known cancer related microRNAs were obtained from miR2Disease \citep{jiang2008mir2disease} and HMDD \citep{li2013hmdd} databases.

Nowadays, it is very difficult or impossible to collect data for non-cancerous microRNA \citep{lee2013microrna}. For a negative data set, we chose microRNAs that exhibited the lowest fold change values as negative controls by analyzing the corresponding expression profile of the breast cancer. We also used the same number of negative controls as positive ones. We downloaded microRNAs expression profile GSE45666 \citep{lee2013microrna} from GEO. 

We computed three lists of p-value for microRNAs according to our method. We arranged lists of p-value by descending order and we calculated AUC for each list to identify the importance of each factor (TF, disease and DEG) for determining which microRNA is associated with breast cancer.

\begin{figure}[!t]
		\centering
		\includegraphics[width=.8 \textwidth]{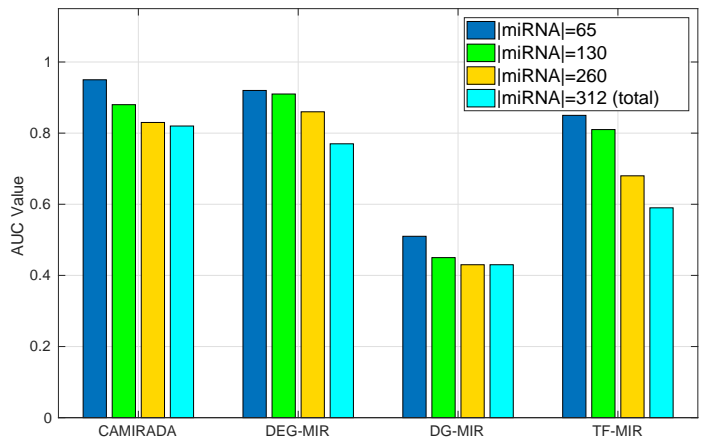}
		\caption{Computing AUC for each three list of p-value for DEG-microRNAs, Disease-microRNAs and TF-microRNAs. And we calculated AUC for CAMIRADA.}\label{fig:2}	
\end{figure}

 As the Figure \ref{fig:2} shows, the lowest AUC is for disease-microRNAs and the maximum AUC is for CAMIRADA approach. So, this shows that DEGs and TFs influence on identification of which microRNAs are related to breast cancers. For using the influence of TFs and DEGs on identification of microRNAs, we combined p-vlues with some tests in R.

The tests included Wilkinsonp, sumlog, sumz, logitp, meanp and sump from metap \citep{MetaP} package in R. Then we scored microRNAs according to our model. For each p-value as threshold, we computed $TP$, $TN$, $FP$ and $FN$ by using the following definition:
\begin{itemize}
\item TP: The number of microRNAs are in positive set and their scores are below the threshold.
\item FP: The number of microRNAs are in negative set or they are not in positive set and their scores are below the threshold.
\item TN: The number of microRNAs are in negative set or they are not in positive and their scores are above the threshold.
\item FN: The number of microRNAs are in positive set and their scores are above the threshold.
\end{itemize}

\begin{figure}[!t]
		\centering
		\includegraphics[width=.9 \textwidth]{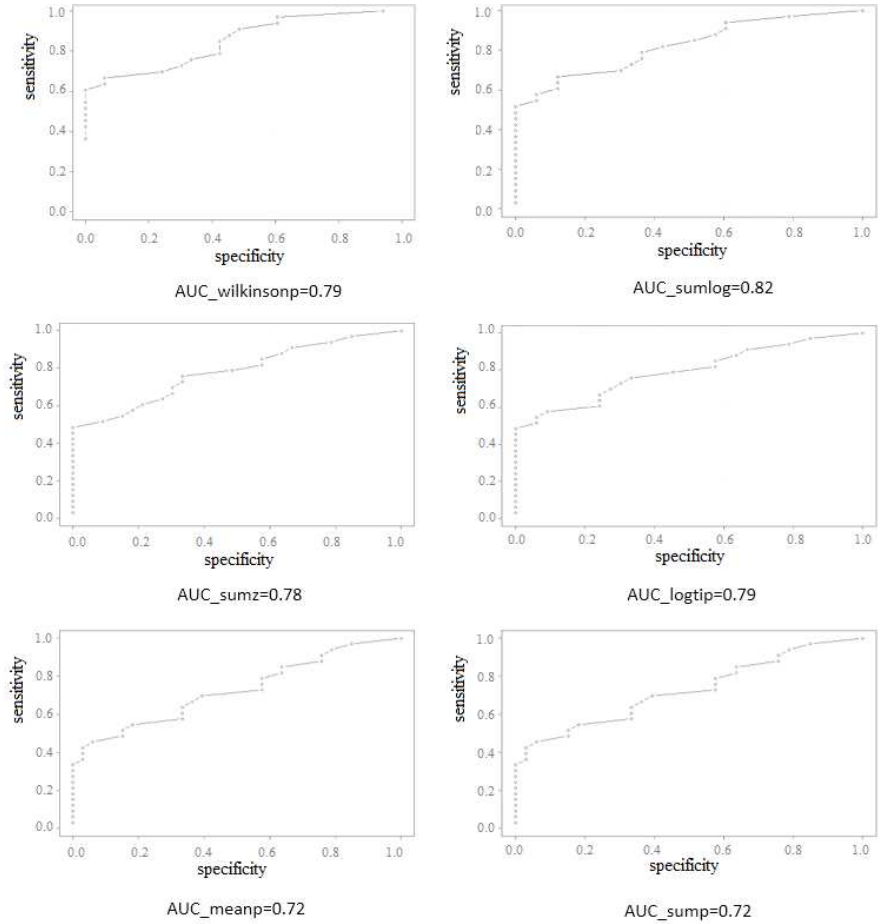}
		\caption{The AUC for all of the p-value combining tests.}\label{fig:3}	
\end{figure}

To select the best test for combining the p-value, we combined the three lists of p-values by all tests and then we calculated AUC for each tests results are shown in Figure \ref{fig:3}. As shown in Figure \ref{fig:3}, the best result is for sumlog test by 82\%, so we choose this test for combining the three lists of p-values.

\begin{figure}[!t]
		\centering
		\includegraphics[width=.8\textwidth]{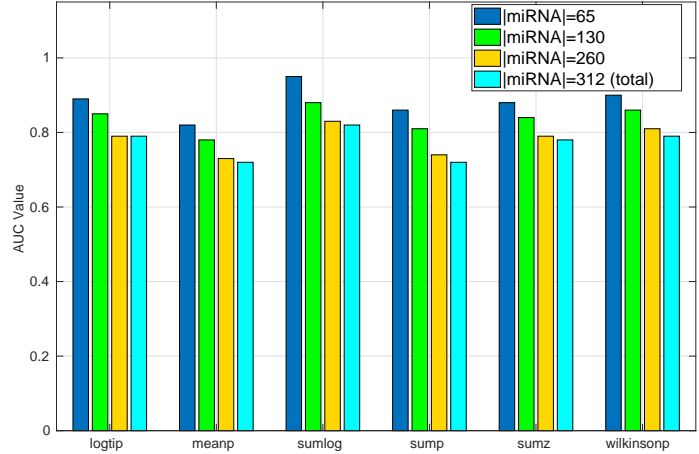}
		\caption{Bar chart for comparing the tests to combine the p-value.}\label{fig:4}	
\end{figure}

For more comparing the tests to combine the p-value we combined the three lists of p-values and we calculated AUC for 65, 130, and 260 number of the microRNAs from the top of the ranked list of microRNAs and the result of the comparison of tests are shown by bar chart in Figure \ref{fig:4}.

According the results that achieved from tests of p-value combining, sumlog choose to combine p-value. So, we combined three ranked list of p-value with sumlog and then arranged the list of combined p-vlaues in ascending order. The lower the p-value for a microRNA, the greater the probability that this microRNA is associated with the cancer. 

In Table \ref{tab:Table 1}  the 10 top of the list of the ranked microRNAs with CAMIRADA and the role of microRNAs in breast cancer are shown.

\begin{table}
	\centering
	\caption{The 10 top of the list of the ranked microRNAs.}
	\label{tab:Table 1}
	\begin{tabular}{lll}
		\toprule
		microRNAs & p-value & role of microRNAs in breast cancer\\
		\midrule
		hsa-miR-93 \citep{shyamasundar2016mir} & 0.000679 & Tumor suppressor \\
		hsa-miR-526b \citep{landman2014role} & 0.000702 & Oncogene \\
		hsa-miR-17 \citep{hossain2006mir} & 0.000951 & Oncogene \\
		hsa-miR-20b \citep{li2012differential} & 0.00125 & Oncogene \\
		hsa-miR-199a \citep{chen2016mir} & 0.001486 & Tumor suppressor \\
		hsa-miR-20a \citep{li2015epigenetic} & 0.001529 & Oncogene \\
		hsa-miR-149 \citep{fang2016down} & 0.0020011 & Oncogene \\
		hsa-miR-199b \citep{lehmann2010identification} & 0.00265 & Biomarker \\
		hsa-miR-519d [38] & 0.002863 & Oncogene \\
		hsa-miR-502 \citep{cheng2012microrna} & 0.036284 & Tumor suppressor\\
		\bottomrule
	\end{tabular}
\end{table}

\section{Conclusion}\label{IV}
The identification of novel cancer-associated microRNAs is important for diagnosing cancer in primary steps for treatment the cancer. In this study, we presented a new approach (CAMIRADA) to identify microRNAs related to breast cancer. CAMIRADA find the relationship between microRNAs and cancer and it ranked microRNAs base on their relation with cancer, by using the microRNAs-disease genes associated, the DEGs-microRNAs related, and the TFs-microRNAs relation. To evaluated CAMIRADA algorithm we calculated AUC which was 0.95 for 65 microRNAs in top of the ranked list of microRNAs. CAMIRADA showed that TFs and DEGs have important roles for identification of which microRNA is related to cancer, so for showing the influence of them we combined three ranked lists of p-values by sumlog test.

By using CAMIRADA, we predicted eight novel microRNAs including hsa-miR-93, hsa-miR-526b, hsa-miR-20b, hsa-miR-199a, hsa-miR-6884, hsa-miR-199b, and hsa-miR-519d associated to breast cancer which are not yet recorded in the disease-microRNA association HMDD and miR2Disease. For future studies, our methods can be used to identify which microRNAs are related which cancers, we can make a network for each microRNA and associated cancers, then analysis the network and extract information to identify related between cancers.

\section*{References}
\bibliographystyle{unsrtnat}
\bibliography{Ref}

\end{document}